\documentclass[twocolumn, tighten, times]{aastex61}
\usepackage{natbib, amsmath, amssymb, graphicx, lineno}

\graphicspath{{figures/}}

\begin{document}

\newcommand{\illus}{\textit{Illustris}}
\newcommand{\sdss}{SDSS}
\newcommand{\sdssfull}{Sloan Digital Sky Survey}
\newcommand{\stelmass}{M_{\star}}
\newcommand{\solmass}{M_{\odot}}
\newcommand{\solmassval}{1.989\times10^{30}\;{\rm kg}}
\newcommand{\solmasswval}{M_{\odot} = 1.989\times10^{30}\;{\rm kg}}
\newcommand{\stellum}{L_{\star}}
\newcommand{\sollum}{L_{\odot}}
\newcommand{\sollumval}{3.828\times10^{26}\;{\rm W}}
\newcommand{\sollumwval}{L_{\odot} = 3.828\times10^{26}\;{\rm W}}

\newcommand{\revision}[3]{#2}


\title{Galaxy Zoo: Morphological classification of galaxy images from the \textit{Illustris} simulation}
\author{Hugh Dickinson}
\affiliation{School of Physics and Astronomy, University of Minnesota, 116 Church Street SE, Minneapolis, MN 55455, USA}

\author{Lucy Fortson}
\affiliation{School of Physics and Astronomy, University of Minnesota, 116 Church Street SE, Minneapolis, MN 55455, USA}

\author{Chris Lintott}
\affiliation{Oxford Astrophysics, University of Oxford, Denys Wilkinson Building, Keble Road, Oxford, OX1 3RH, UK}

\author{Claudia Scarlata}
\affiliation{School of Physics and Astronomy, University of Minnesota, 116 Church Street SE, Minneapolis, MN 55455, USA}

\author{Kyle Willett}
\affiliation{School of Physics and Astronomy, University of Minnesota, 116 Church Street SE, Minneapolis, MN 55455, USA}

\author{Steven Bamford}
\affiliation{School of Physics and Astronomy, The University of Nottingham, University Park, Nottingham, NG7 2RD, UK}

\author{Melanie Beck}
\affiliation{School of Physics and Astronomy, University of Minnesota, 116 Church Street SE, Minneapolis, MN 55455, USA}

\author{Carolin Cardamone}
\affiliation{Department of Mathematics and Science, Wheelock College, Boston, MA 02215, USA}

\author{Melanie Galloway}
\affiliation{School of Physics and Astronomy, University of Minnesota, 116 Church Street SE, Minneapolis, MN 55455, USA}

\author{Brooke Simmons}
\altaffiliation{Einstein Fellow}
\affiliation{Center for Astrophysics and Space Sciences, Department of Physics, University of California, San Diego, CA 92093, USA}

\author{William Keel}
\affiliation{Department of Physics and Astronomy, University of Alabama, Box 870324, Tuscaloosa, AL 35487, USA}

\author{Sandor Kruk}
\affiliation{Oxford Astrophysics, University of Oxford, Denys Wilkinson Building, Keble Road, Oxford, OX1 3RH, UK}

\author{Karen Masters}
\affiliation{Institute for Cosmology and Gravitation, University of Portsmouth, Dennis Sciama Building, Burnaby Road, Portsmouth, PO1 3FX, UK}

\author{Mark Vogelsberger}
\affiliation{Department of Physics, Kavli Institute for Astrophysics and Space Research, Massachusetts Institute of Technology, Cambridge, MA 02139, USA}

\author{Paul Torrey}
\altaffiliation{Hubble Fellow}
\affiliation{Department of Physics, Kavli Institute for Astrophysics and Space Research, Massachusetts Institute of Technology, Cambridge, MA 02139, USA}

\author{Gregory {F.} Snyder}
\affiliation{Space Telescope Science Institute, 3700 San Martin Drive, Baltimore, MD 21218, USA}

\correspondingauthor{Hugh Dickinson}

\begin{abstract}
 Modern large-scale cosmological simulations model the universe with increasing sophistication and at higher spatial and temporal resolutions. These ongoing enhancements permit increasingly detailed comparisons between the simulation outputs and real observational data. Recent projects such as \illus\ are capable of producing simulated images that are designed to be comparable to those obtained from local surveys. This paper tests the degree to which \illus\ achieves this goal across a diverse population of galaxies using visual morphologies derived from Galaxy Zoo citizen scientists. Morphological classifications provided by these volunteers for simulated galaxies are compared with similar data for a compatible sample of images drawn from the \sdssfull\ (\sdss)\ Legacy Survey. This paper investigates how simple morphological characterization by human volunteers asked to distinguish smooth from featured systems differs between simulated and real galaxy images. Significant differences are identified, which are most likely due to the limited resolution of the simulation, but which could be revealing real differences in the dynamical evolution of populations of galaxies in the real and model universes. Specifically, for stellar masses $M_{\star}\lesssim10^{11} M_{\odot}$, a substantially larger proportion of \illus\ galaxies that exhibit disk-like morphology or visible substructure, relative to their \sdss\ counterparts. Toward higher masses, the visual morphologies for simulated and observed galaxies converge and exhibit similar distributions. The stellar mass threshold indicated by this divergent behavior confirms recent works using parametric measures of morphology from \illus\ simulated images. When $M_{\star}\gtrsim10^{11} M_{\odot}$, the \illus\ dataset contains substantially fewer galaxies that classifiers regard as unambiguously featured. In combination, these results suggest that comparison between the detailed properties of observed and simulated galaxies, even when limited to reasonably massive systems, may be misleading.

\end{abstract}

\section{Introduction}
As large-scale simulations of the universe increase in size and in resolution, increasingly sophisticated comparisons with observations are becoming more feasible. While early work concentrated on matching features of the universe captured by simple parameterizations such as the mass function or scaling relations \citep[e.g.][]{1993MNRAS.264..201K,1994MNRAS.271..781C}, modern cosmological simulations produce galaxies with apparently realistic star formation histories, substructures, and colors \citep[e.g.][]{2014MNRAS.445..175G,2015MNRAS.450.1937C,2017MNRAS.467.4739K}. The prospect of ``observing'' this simulated universe via the creation of artificial images offers the chance to test any such simulation's fidelity, and any discrepancies may provide new insights on the physics that drives galaxy formation and evolution.

The obvious comparison for simulations that model the present-day galaxy population is the \sdssfull\ (\sdss; \citet{2000AJ....120.1579Y,2002AJ....124.1810S}), which has provided a wealth of information about a large number of local systems \citep[see][for just some of the most highly cited results]{2001AJ....122.1861S,2003MNRAS.346.1055K, 2004ApJ...613..898T, 2004MNRAS.351.1151B, 2004ApJ...600..681B}. The \sdss\ augments its galaxy catalogs with a rich suite of spectral, photometric, and instrumental metadata. In particular, the availability of estimated galaxy redshifts and stellar masses is critical for our analysis.

Modern simulations such as \illus\ \citep{2014Natur.509..177V, 2014MNRAS.444.1518V, 2014MNRAS.445..175G, 2015MNRAS.452..575S} have been used to construct simulated versions of the \sdss\ \citep{2015MNRAS.447.2753T}, and comparisons between observed and simulated universes have utilized a large range of parameters derived from observations \citep{2015MNRAS.454.1886S, 2017MNRAS.tmp...30B, 2017MNRAS.467.2879B}. However, much insight can still be gained by relying on morphological classification of galaxy images. Morphology is a sensitive probe of a galaxy's dynamical and star formation histories, and such classifications have been shown to reflect differences between systems that are often difficult to recover from purely parametric approaches \citep[e.g.][]{2009MNRAS.393.1324B,2009MNRAS.396..818S, 2010MNRAS.405..783M}, and have also helped to unveil previously unnoticed trends and behaviors \citep[e.g.][]{2010ApJ...711..284S, 2011MNRAS.411.2026M, 2013MNRAS.429.2199S, 2013MNRAS.429.1051C, 2015MNRAS.448.3442G, 2016MNRAS.463.2986S,2014MNRAS.440.2944K}.

This paper uses visual morphological classifications as a metric for comparison between simulated and observed universes. Using calibrated citizen science data from the Galaxy Zoo project \citep{2008MNRAS.389.1179L, 2013MNRAS.435.2835W}, we provide non-parametric labels for a large number of simulated galaxies and compare these to \sdss\ galaxies labeled in the same way. In this manner, we aim to investigate the degree to which large cosmological simulations, and specifically \illus, can claim to match the present-day galaxy population.

\newpage
\section{Data}
\subsection{The \illus\ Sample}\label{sec:illus_sample}
\illus\ is a suite of large volume, cosmological hydrodynamical simulations run with the moving-mesh code Arepo \citep{2010MNRAS.401..791S, 2014MNRAS.445..175G}. It includes a comprehensive set of physical models that are deemed critical for modeling the formation and evolution of galaxies across cosmic time. Galaxy formation processes in \illus\ are simulated following the models described by \citet{2013MNRAS.436.3031V} and \citet{2014MNRAS.438.1985T}. Each of the \illus\ simulations encompasses a volume of $106.5\;{\rm Mpc}^{3}$ and self-consistently evolves five different types of resolution element (dark matter particles, gas cells, passive gas tracers, particles that represent stars and their stellar winds, and supermassive black holes) from a starting redshift of $z=127$ to the present day, $z=0$. The \illus\ simulation suite successfully reproduces a range of well established galaxy scaling relations. It implements a unique combination of high-resolution and total simulation volume, which provides an ideal test dataset for our purposes.

The \illus\ image sample is generated using an ensemble of 6891 unique subhaloes that had assembled within the \illus\ simulation volume by $z=0$. Each subhalo is assumed to represent a single galaxy. These were chosen to have $\stelmass \gtrsim 10^{10} \solmass$\footnote{\illus\ generates several definitions of the stellar mass for each simulated galaxy. Throughout this paper, we use the \textit{total} stellar mass, labeled as \texttt{mass\_stars} in the \illus\ catalog.}, which corresponds to a typical number of stellar particles $\gtrsim 10^{5}$. Simulated galaxies comprised of fewer particles were deemed unlikely to accurately represent morphological features of interest \citep[e.g.][]{2015MNRAS.447.2753T}, and were therefore excluded from our sample.

We use images from \cite{2015MNRAS.447.2753T}, which have been processed as described in \cite{2015MNRAS.454.1886S} to produce `observationally realistic' images. This process produces synthetic \illus\ images that are square arrays with side length 424 pixels, with a typical angular pixel scale $0\farcs05-0\farcs10$ per pixel. For each image, the precise pixel scaling is adjusted to ensure that the central $\frac{2}{3}$ of each subject image corresponds to twice the simulated galaxy's projected Petrosian radius. This scaling emulates the approach used to generate the original Galaxy Zoo 2 subject images. Each image is convolved with a nominal PSF with Full Width at Half Maximum (FWHM) $\sim1\farcs0$, which is similar to the $\sim1\farcs4$ average seeing for the \sdss\ DR7; the two sets of images should be broadly comparable. It should be noted that these images represent a simulation of galaxies that have evolved until redshift zero, but projected as if they lie at $z=0.05$. We expect little evolution in the galaxy population between $z=0.05$ and the present, and so this displacement should not significantly affect the comparison we wish to make. Observational evidence also indicates that galaxy populations in the real universe exhibit little evolution in this redshift interval \citep[e.g.][]{2003ApJ...599..847R, 2003ApJ...594..186B}.

Images of each galaxy were generated for four orientations that model observation from the separate vertices of a tetrahedron with the subhalo at the center (the tetrahedron is oriented with respect to the simulation and so randomly relative to the galaxy). Backgrounds are randomly selected from real \sdss\ images. The `target' galaxy is assumed to be in the foreground and in rare cases may be superimposed over systems that are actually closer than the projected distance of the simulated galaxy ($z=0.05$). Four separate backgrounds for each galaxy were used to mitigate this and other systematic effects. The final sample that is potentially available for classification therefore comprises a total of 16 images per subhalo, making a total of 110,256 distinct subjects.

\subsection{The \sdss\ Sample}\label{sec:sdss_sample}
To provide a valid comparison for the \illus\ sample, described in $\S$\ref{sec:illus_sample}, we begin by selecting \sdss\ galaxies with $\stelmass > 10^{10}\;\solmass$ and with redshifts between $z=0.045$ and $z=0.055$.

\begin{figure*}
 \begin{center}
  \includegraphics[width=0.45\textwidth]{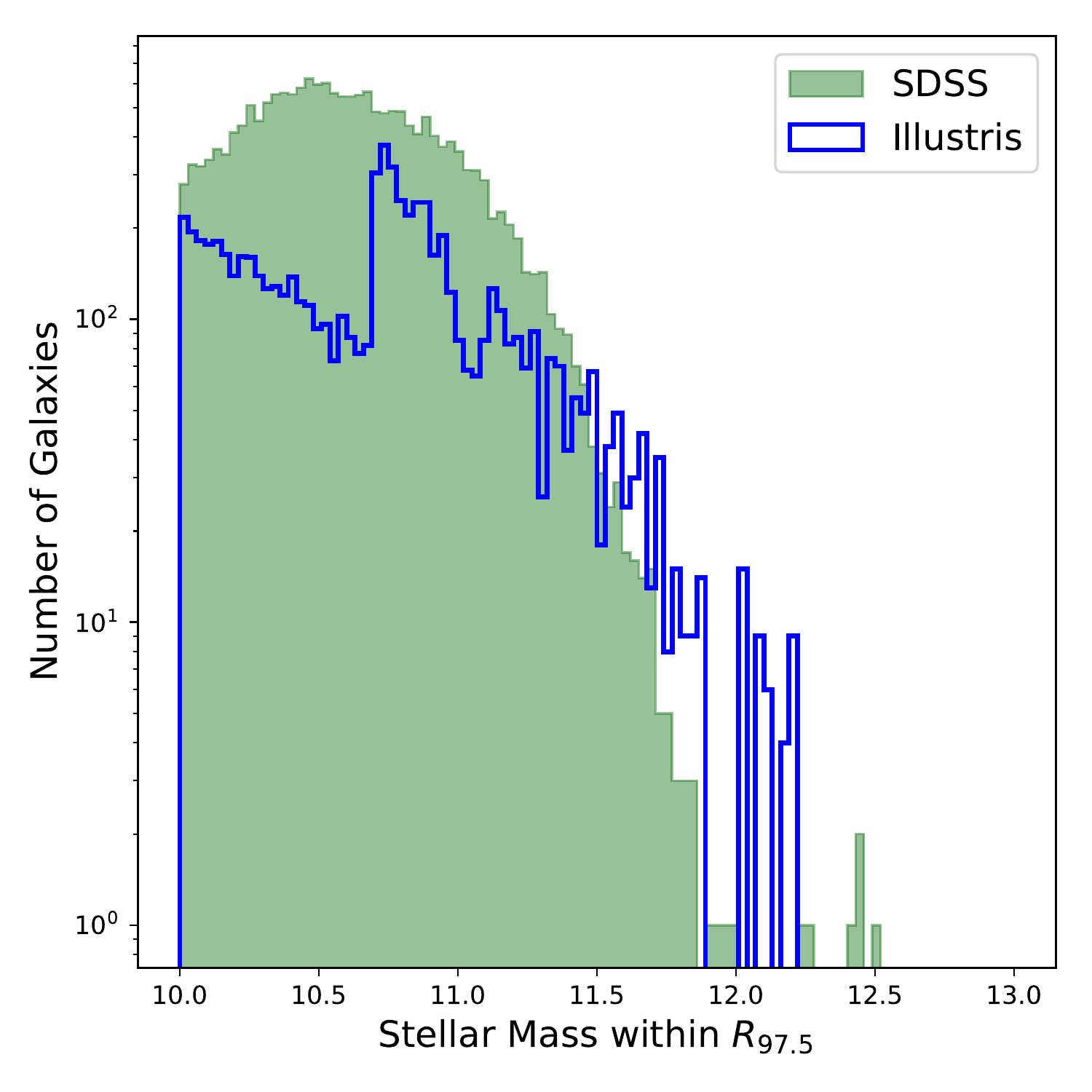}
  \includegraphics[width=0.45\textwidth]{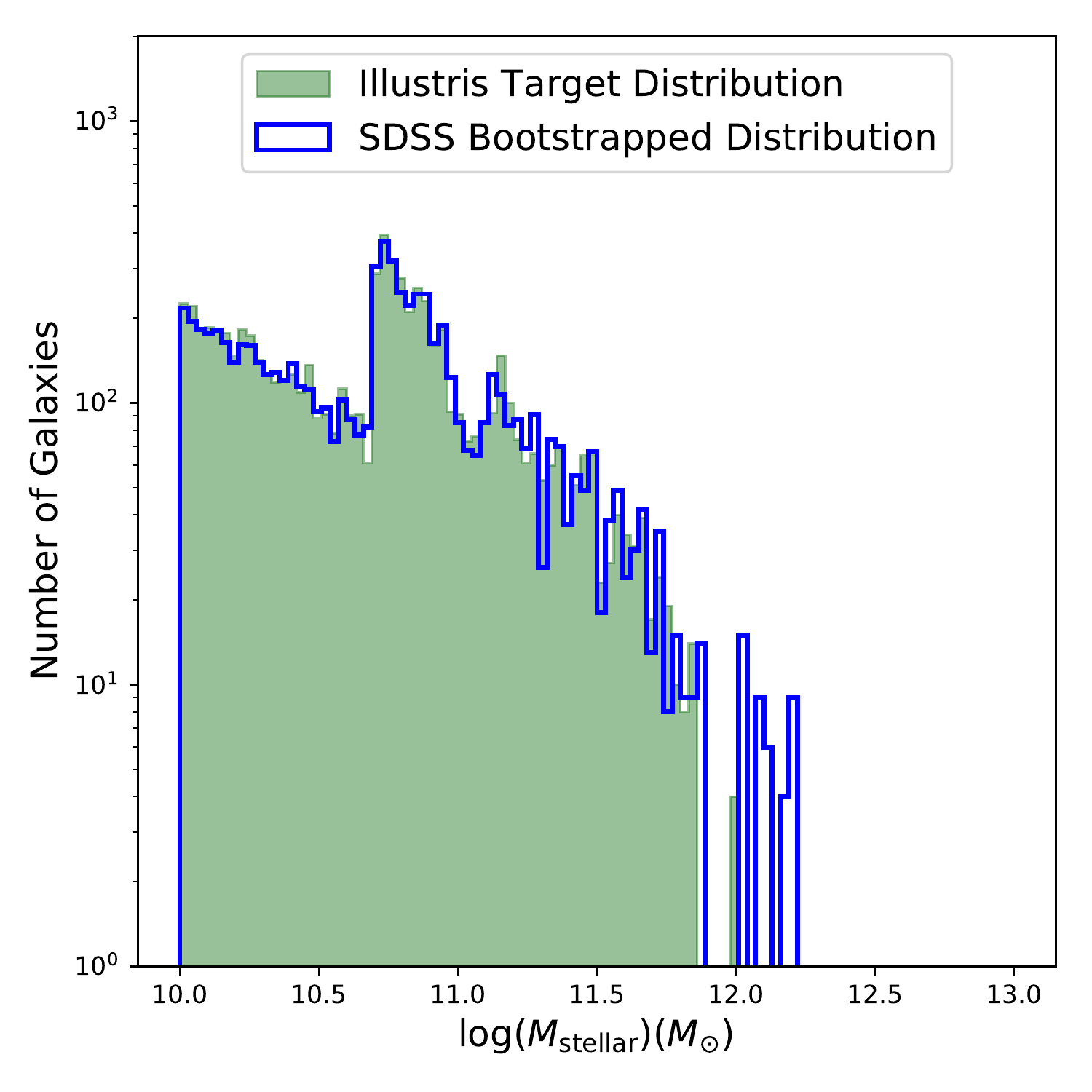}
  \caption{Raw (left) and resampled (right) stellar mass distributions for the \illus\ (blue hollow) and \sdss\ (green filled) datasets. Distributions are shown for the inferred stellar mass within 97.5\% (left) of the galaxy's Petrosian radius.}\label{fig:sampleMassDists}
 \end{center}
\end{figure*}

The left-hand panel of Figure \ref{fig:sampleMassDists} shows the stellar mass\footnote{Stellar masses for the \sdss\ galaxies were derived from the \texttt{P97P5} column of the MPA-JHU catalog \citep{2004MNRAS.351.1151B}.} distributions of the raw, redshift-selected \sdss\ and \illus\ datasets. The distributions are obviously mismatched due to a combination of the a-priori galaxy mass selection applied to the \illus\ sample and incomplete sampling of faint, low-mass galaxies in the \sdss.

Within the narrow redshift range spanned by our \sdss\ sample, the inferred stellar mass provides a good proxy for galactic size and luminosity, which are both likely to influence the observability of morphological features. We therefore use bootstrap resampling to construct a final \sdss\ sample with a mass distribution that matches the \illus\ sample \revision{}{that was ultimately classified (see $\S$\ref{sec:gz_classification_infrastructure})}{dickinson}. The \sdss\ sample is drawn from 100 bins, equally separated in log-mass space. The right-hand panel of Figure \ref{fig:sampleMassDists} illustrates the resulting distribution in $M_{\star}$ of our bootstrap-resampled \sdss\ dataset. This dataset contains 7159 entries, of which 5556 are unique. Among those remaining images that are sampled repeatedly, the vast majority are pairs; very few images appear more than twice.

\begin{figure*}[htb!]
 \begin{center}
  \includegraphics[width=0.99\textwidth]{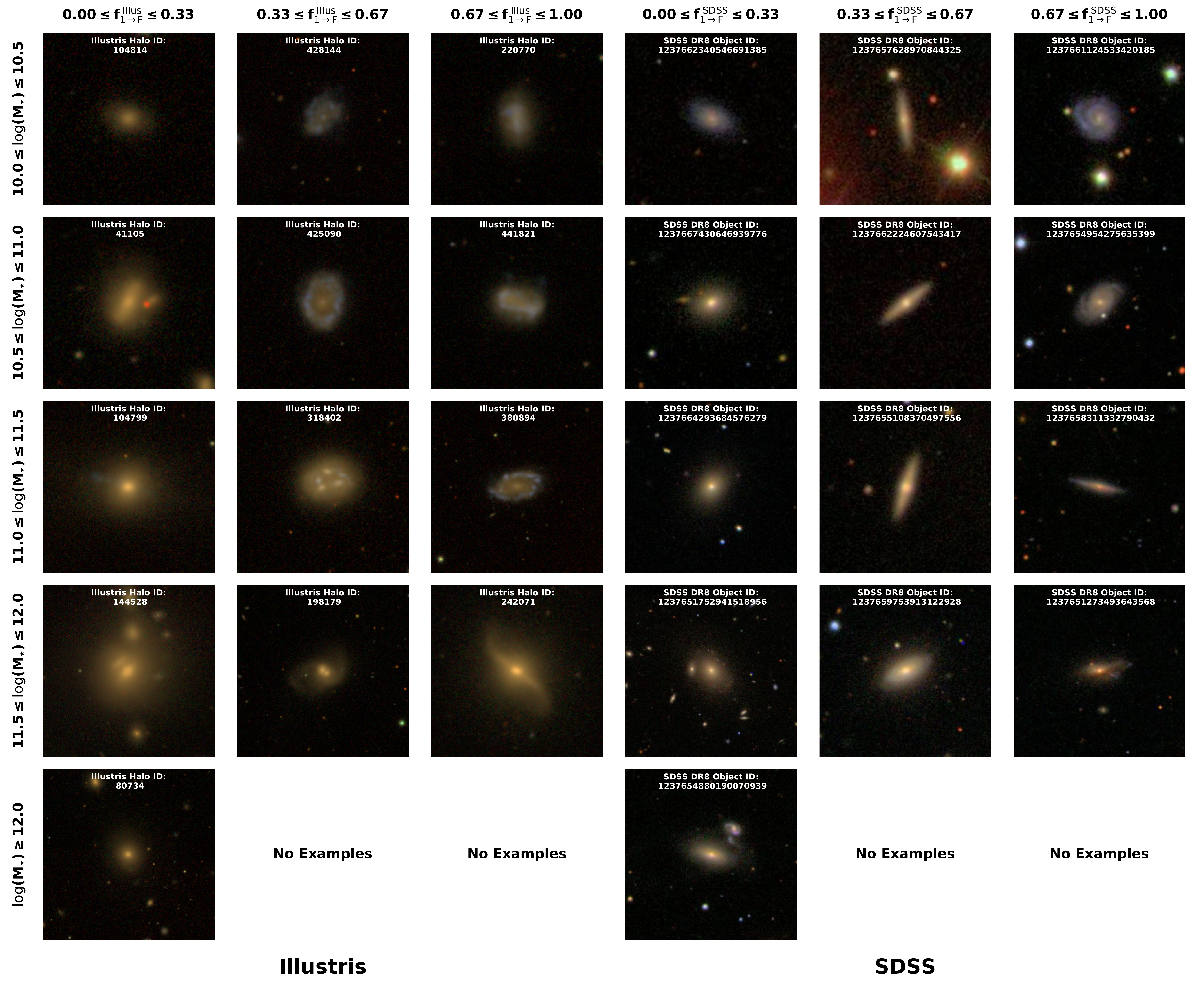}
  \caption{Comparison between mass-matched \illus\ (three left-hand columns) and \sdss\ (three right-hand columns) subject images. Each row shows a triplet of galaxies drawn from broad mass bins for each survey. Listing from the top row to the bottom, the chosen mass bins correspond to $10\leq\log(M_{\star}/M_{\odot})\leq 10.5$, $10.5\leq\log(M_{\star}/M_{\odot})\leq 11.0$, $11.0\leq\log(M_{\star}/M_{\odot})\leq 11.5$, $11.5\leq\log(M_{\star}/M_{\odot})\leq 12$, and $12\leq\log(M_{\star}/M_{\odot})\leq 50$}
  \label{fig:sdssIllustrisSubjectComparison}
 \end{center}
\end{figure*}

For reference, Figure \ref{fig:sdssIllustrisSubjectComparison} compares mass-matched, but otherwise randomly selected images from the \illus\ and \sdss\ subject sets.

\subsection{Predictable differences between the \illus\ and \sdss\ images}\label{subsec:remaining_inconsistencies}
\begin{figure*}[htb!]
 \begin{center}
  \includegraphics[width=0.99\textwidth]{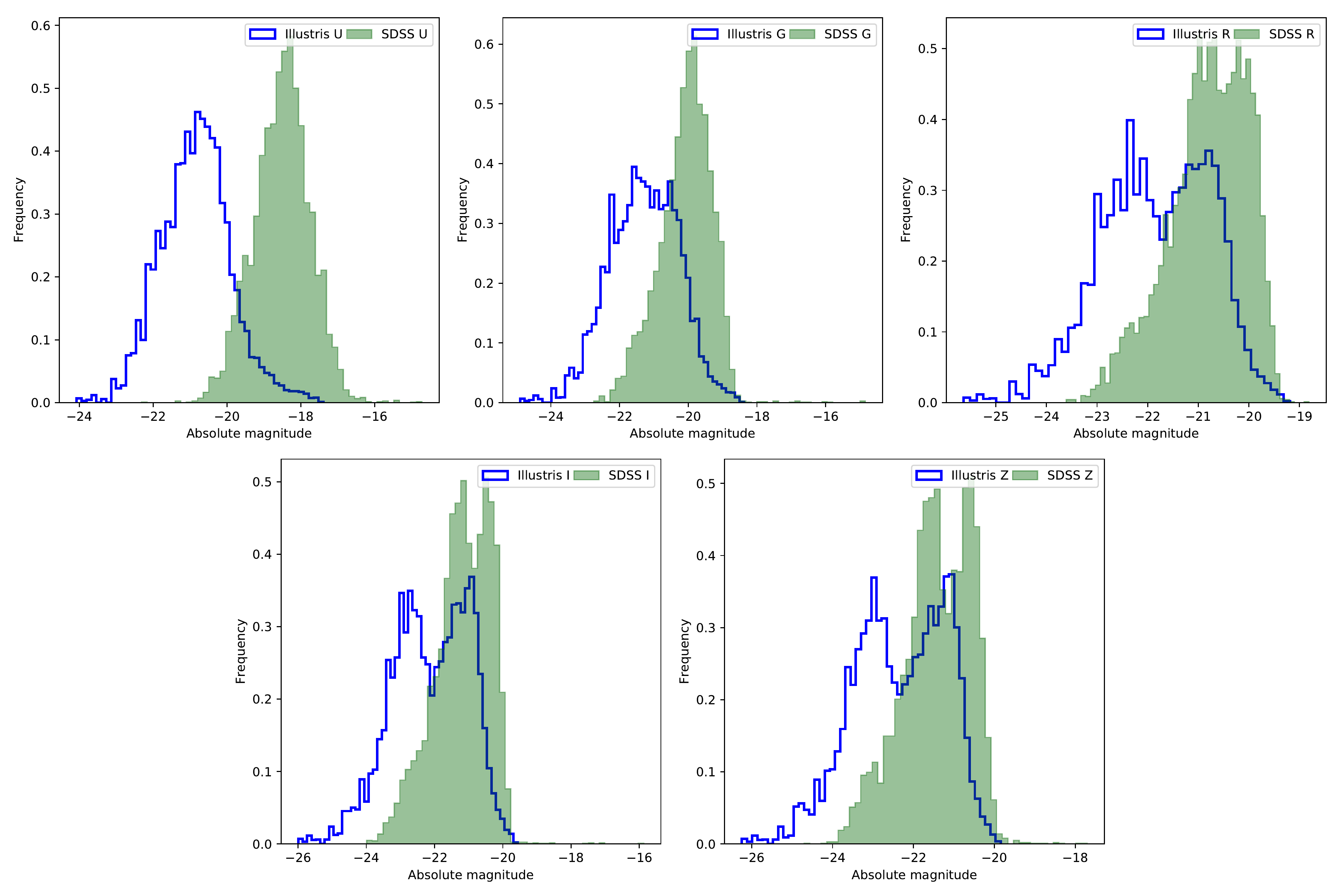}
  \caption{Illustration of the mismatch between the distributions of absolute magnitude for the 5 \sdss\ filters ($u, g, r, i, z$) for the \illus\ and \textit{resampled} \sdss\ datasets.}
  \label{fig:sdssBandMagDists}
 \end{center}
\end{figure*}

Several assumptions and simplifications were adopted when generating synthetic galaxy images based on the \illus\ simulation data. Accordingly, some predictable differences between simulated and real images are inevitable, and we outline the most significant of these here. Intrinsic dust reddening was not considered when generating synthetic images based upon the simulated \illus\ galaxy structures. Dust formation occurs in dense molecular clouds, which are not fully resolved at the $\sim 1\;\mathrm{kpc}$ spatial resolution that \illus\ achieves, so modeling of the dust within simulated galaxies requires augmentation of the simulation output with a number of \textit{ad-hoc} assumptions\footnote{\revision{}{\citet{2015MNRAS.452.2879T} showed how different modeling assumptions pertaining to dust obscuration affect the inferred observational colours of simulated galaxies in the EAGLE simulation.}{dickinson}}. In contrast, the three-dimensional positions of the \illus\ galaxies' stellar populations are directly resolved by the simulation. Accordingly, synthetic images that omit dust modeling provide a faithful representation of the raw simulation output, which ultimately simplifies inference of the performance of \illus\ using visual classification data. Nonetheless, dust obscuration is known to be significant for some local galaxies \citep[e.g.][]{2010MNRAS.404..792M}, and this omission is manifested in Figure \ref{fig:sdssBandMagDists} as clear mismatches between the distributions of absolute magnitude for the five \sdss\ filters ($u, g, r, i, z$) between the \illus\ and \textit{resampled} \sdss\ samples that worsens for increasingly blue filters.

In addition, \cite{2015MNRAS.454.1886S} note that the sizes of simulated and real galaxies (measured by the half-mass radius for \illus\ and Petrosian 50\% radius for \sdss) are comparable at masses of $10^{11}$ and above, but at lower $\stelmass$ the \illus\ galaxies are comparatively more extended. The discrepancy amounts to a factor of two at a mass of $10^{10} \solmass$.

\section{Galaxy Zoo Classification Infrastructure}\label{sec:gz_classification_infrastructure}
Galaxy Zoo is a set of citizen science projects that have collectively engaged hundreds of thousands of volunteers in the classification of galaxy images drawn from large ground-based surveys and from those conducted by the \emph{Hubble Space Telescope} \citep{2008MNRAS.389.1179L, 2012amld.book..213F}. Such classifications have been shown to be a good match to expert classifications \citep{2008MNRAS.389.1179L, 2013MNRAS.435.2835W, 2017MNRAS.464.4420S, 2017MNRAS.464.4176W}. Moreover, the degree of consistency between the classifications provided by multiple volunteers for the same galaxy image provides a measure of the precision of their aggregate classification.

Classification of a galaxy image in Galaxy Zoo entails answering a series of questions, each
evaluating a particular aspect of a galaxy's morphological appearance. The earliest questions segregate the subject set into broad morphological categories before subsequent questions investigate increasingly intricate aspects of a galaxy's appearance. The full question set is subjected to hierarchical filtering such that questions are only asked if they remain pertinent following earlier responses. Accordingly, sampling becomes increasingly sparse for questions that appear later in the classification hierarchy and the degree of statistical uncertainty associated with each subject's consensus response increases. For this project, \illus\ images were classified via a decision tree emulating the tree used for the Galaxy Zoo 2 project and described in \citet{2013MNRAS.435.2835W}.

In Galaxy Zoo, each galaxy image is classified by at least forty\footnote{The mean number of classifications per subject is 40.2.} nominally independent volunteers. The individual responses to each question are then aggregated to yield an overall consensus classification. For questions that require a binary response, the availability of multiple independent responses permits the aggregate classification to be encapsulated as a\ real-valued vote fraction, which is evaluated as the ratio of the number of positive (or negative) responses to the total number of responses.

The \illus\ classifications used for this study were accumulated via the Galaxy Zoo web-based interface between 2015 September and 2017 August. During this interval, 164,627 volunteers contributed 814,283 morphological assessments for 20248 distinct galaxy images. Classification began with an initial subject set comprising 17046 images for simulated galaxies with stellar masses $10\leq\log_{10}(M_{\star}/M_{\odot})\leq13$ \revision{}{The initial sample was designed to facilitate the assessment of potential systematic biases that were anticipated but were not ultimately evident during analysis. To isolate the effect of background and viewing angle on morphological classification, a subset of 10832 images were derived from 677 distinct subhaloes that were selected by uniform random sampling from within two narrow ranges of total \textbf{halo} mass $10.5\leq\log_{10}(M_{\mathrm{halo}}/M_{\odot})\leq11,\,12.5\leq\log_{10}(M_{\mathrm{halo}}/M_{\odot})\leq13$. Each subhalo was imaged from the four directions corresponding with the vertices of a regular tetrahedron and superimposed over four randomly selected background images per vertex, as described in $\S$\ref{sec:illus_sample}. The remaining 6214 images sample the complementary ranges of halo mass, facilitating mass-independent morphological comparison with observed \sdss\ galaxies. Each synthetic image in this subset corresponds to a distinct subhalo, viewed from a single, randomly selected viewing angle and superimposed over a single randomly selected background.}{dickinson}. \revision{This}{To enhance the sample of classifications for the most massive \illus\ galaxies, the initial}{dickinson} set was subsequently augmented with 3202 additional images for which the corresponding stellar masses exceeded $10^{10.5}M_{\odot}$.

For our \sdss\ sample, we use data from Galaxy Zoo 2 \citep{2013MNRAS.435.2835W}, which provides detailed morphological classifications of nearly 250,000 galaxies drawn from the 7th \sdss\ data release \citep{2009ApJS..182..543A}. The subset of the \sdss\ used for Galaxy Zoo 2 is described by \citet{2013MNRAS.435.2835W} and was further subsampled to provide a comparison dataset for the \illus\ images and their corresponding morphologies.

\section{Results}\label{sec:results}
We identify discrepancies between the Galaxy Zoo classifications that were obtained for the \illus\ dataset and those obtained for a redshift- and mass-matched sample of \sdss\ galaxies by comparing the distributions of vote fractions obtained for each sample. For this  investigation, we concentrate on the first, most fundamental question in the Galaxy Zoo 2 decision tree, which distinguished galaxies with features - predominately disk-dominated systems - from those where no such features are apparent. Even this crude distinction reflects significant differences in the underlying dynamical and star formation history of a galaxy, which dictate its visual morphology. Accordingly, it is an excellent test of the realism of the images produced by the \illus\ simulation.

\begin{figure*}
 \begin{center}
  \includegraphics[width=0.5\textwidth]{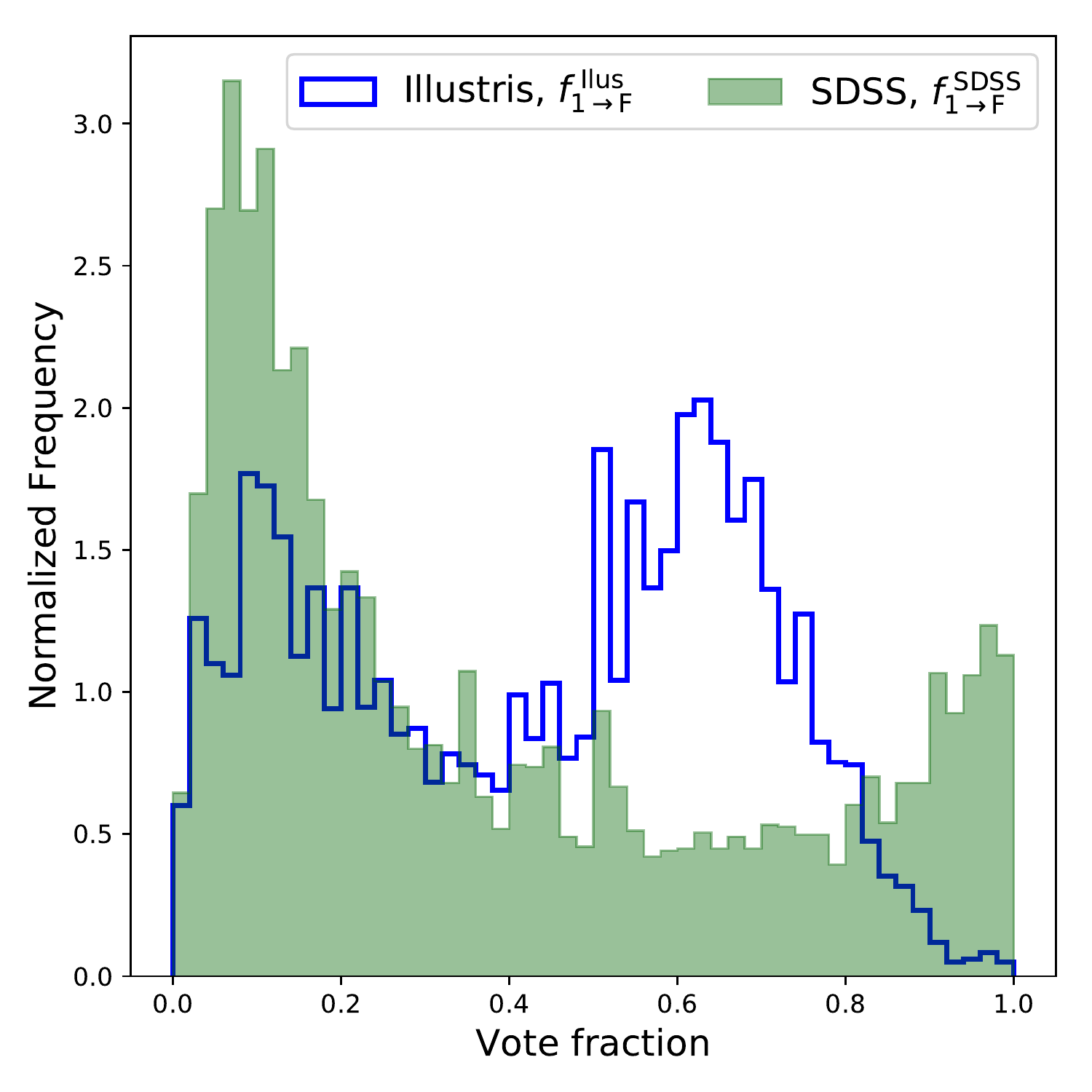}
  \caption{Comparison between the normalized distributions for $f_{\rm 1\rightarrow F}^{\rm Illus}$ and $f_{\rm 1\rightarrow F}^{\rm SDSS}$ corresponding to the full \illus\ and \sdss\ samples, respectively. A high value of $f_{\rm 1\rightarrow F}$ implies that the majority of volunteers discerned discrete substructure in the galaxy image, while $f_{\rm 1\rightarrow F}\rightarrow0$ implies the converse. While the \sdss\ distribution is dominated by systems with low $f_{\rm 1\rightarrow F}^{\rm SDSS}$, the Illustris sample apparently contains many more galaxies that exhibit visible substructure and yield more intermediate vote fractions.}
  \label{fig:featuresVoteFracComparison}
 \end{center}
\end{figure*}

Figure \ref{fig:featuresVoteFracComparison} illustrates the unweighted\footnote{Previous analysis of Galaxy Zoo 2 has used a weighting system, which downweights highly inconsistent classifications; as the population of classifiers has changed between the original GZ2 run and classifications of \illus\ simulated galaxies, introducing such a weighting here would introduce a new systematic difference between the samples. For most systems, the weighting makes little difference in practice. Therefore, we choose to use unweighted vote fractions to avoid even the possibility of introducing a systematic difference between the samples.} vote fraction distributions for the response ``disk or features'' to the question
``Is the galaxy simply smooth and rounded, with no sign of a disk?''
(hereafter $f_{\rm 1\rightarrow F}$) for the \illus\ and \sdss\ samples\footnote{In addition to the nominal positive and negative responses, a third option, which labels the putative galaxy as an ``artifact'' is also possible. All votes for ``artifact'' were discarded when computing the vote fractions we present in this paper. We verified that omitting artifact votes from our analysis does not qualitatively affect our results.}. Consequently, a high value of $f_{\rm 1\rightarrow F}$ implies that the imaged galaxy probably has features, while $f_{\rm 1\rightarrow F}\rightarrow0$ implies the converse. A surprisingly marked disparity is evident. The \sdss\ galaxies show a broadly bimodal distribution, with many (visibly featureless) systems clustered around low featured vote fractions, and a smaller number of systems that have high vote fractions. The \sdss\ distribution arises primarily from genuine morphological separation between elliptical and spiral systems but is augmented at low $f_{\rm 1\rightarrow F}^{\mathrm{SDSS}}$ by galaxies that would exhibit features but are too faint for any intrinsic substructure to be visible in subject images.

The \illus\ sample, by contrast, is characterized by a prevalence of galaxies with visible substructure, which is evident in Figure \ref{fig:featuresVoteFracComparison} as a dominant peak around a modal vote fraction of around 0.6. It is clear from even this simple comparison that there are significant differences between the two samples.

\begin{figure*}[htb!]
 \begin{center}
  \includegraphics[width=\textwidth]{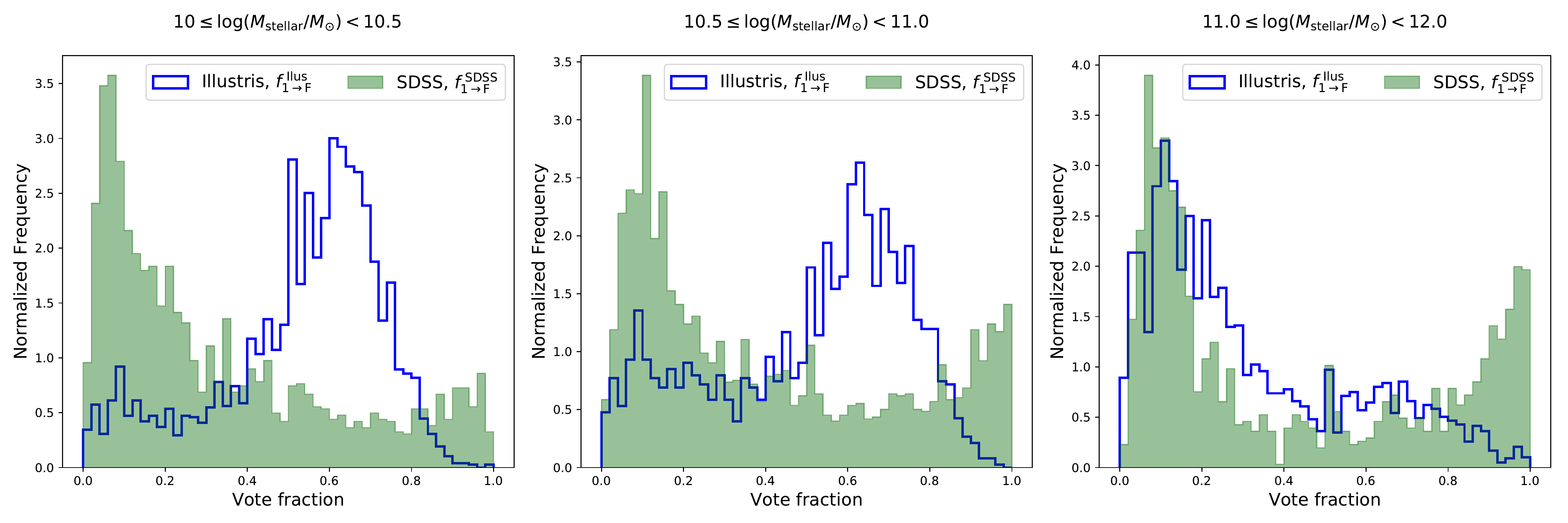}
  \caption{The $f_{\rm 1\rightarrow F}$ vote fractions in intervals of $\log(M_{\star}/M_{\odot})$. Proper interpretation of $f_{\rm 1\rightarrow F}$ is explained in the main text as well as in the caption of Figure \ref{fig:featuresVoteFracComparison}. Below $\log(M_{\star}/M_{\odot})\sim11$, the \sdss\ and \illus\ $f_{\rm 1\rightarrow F}$ distributions match very poorly. At higher masses, overall agreement between the distributions is substantially improved, albeit with a residual discrepancy between the numbers of obviously featured galaxies.}
  \label{fig:featuresVoteFracMassIntervalComparison}
 \end{center}
\end{figure*}

In Figure \ref{fig:featuresVoteFracMassIntervalComparison} we subdivide the \illus\ and \sdss\ samples into disjoint subsamples according to galaxy stellar mass, $M_{\star}$. For $10 \leq \log(M_{\star}/M_{\odot}) \leq 10.5$, the mismatch between the distributions of $f_{\rm 1\rightarrow F}$ that was evident for the full range of galaxy masses is qualitatively reproduced. For subsamples that correspond to higher stellar masses, the $f_{\rm 1\rightarrow F}$ distributions become increasingly similar, and for $M_{\star} \gtrsim 10^{11}\mathrm{M}_{\odot}$, we see a significant fraction of galaxies in the \illus\ sample with low vote fractions as expected from \sdss\ observations.

We verified that the observed overabundance of featured galaxies in \illus\ is not an artifact of viewing angle by individually analyzing four subsets of images corresponding to the distinct vertices of the tetrahedral imaging structure described in $\S$\ref{sec:illus_sample} and verifying that qualitatively similar vote fraction distributions are obtained. We also verified that the observed dependence on $M_{\star}$ is preserved for each subset of the data.

\begin{figure*}[htb!]
 \begin{center}
  \includegraphics[width=0.99\textwidth]{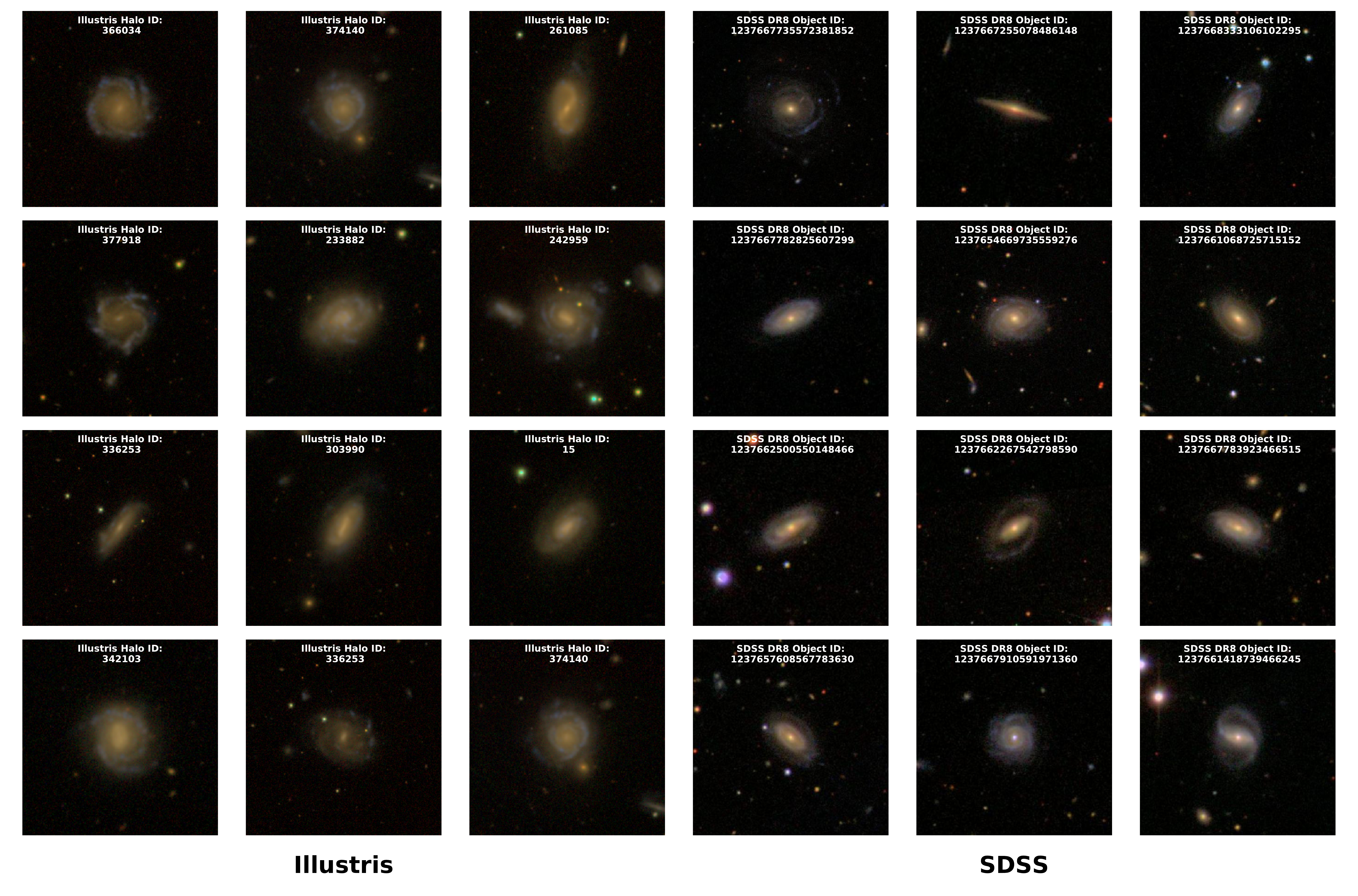}
  \caption{Example images of \illus\ ({three left-hand columns}) and \sdss\ ({three right-hand columns}) galaxies with $f_{\rm 1\rightarrow F} > 0.85$ and $\log(M_{\star}/M_{\odot})>11$}
  \label{fig:highMassHighPFeaturesSDSSAndIllustris}
 \end{center}
\end{figure*}

The other notable difference between the two samples is manifested for $M_{\star} \geq 10^{10.5}\mathrm{M}_{\odot}$ as a significant subset of \sdss\ galaxies with very high featured vote fractions ($f_{\rm 1\rightarrow F} \gtrsim 0.85$). A population of galaxies that almost all classifiers identify as spiral in the \sdss\ is either missing in the simulated universe or classified differently in the \illus\ sample. Figure \ref{fig:highMassHighPFeaturesSDSSAndIllustris} shows representative samples of galaxy images drawn from the mismatching region of ($M_{\star} - f_{\rm 1\rightarrow F})$ parameter space for the \illus\ ({left-hand columns}) and \sdss\ ({right-hand columns}) datasets. While \illus\ does produce a population of featured galaxies with $M_{\star} \geq 10^{10.5}\mathrm{M}_{\odot}$, the \sdss\ image sample appears to include a larger fraction of nearby grand design spirals that the majority of volunteers would classify as obviously featured. In contrast, the \illus\ galaxy images appear slightly more ambiguous, with less prominent disks, and it seems plausible that the apparent deficiency of galaxies that are unanimously perceived as featured reflects this ambiguity.

\begin{figure*}[htb!]
 \begin{center}
  \includegraphics[width=0.9\textwidth]{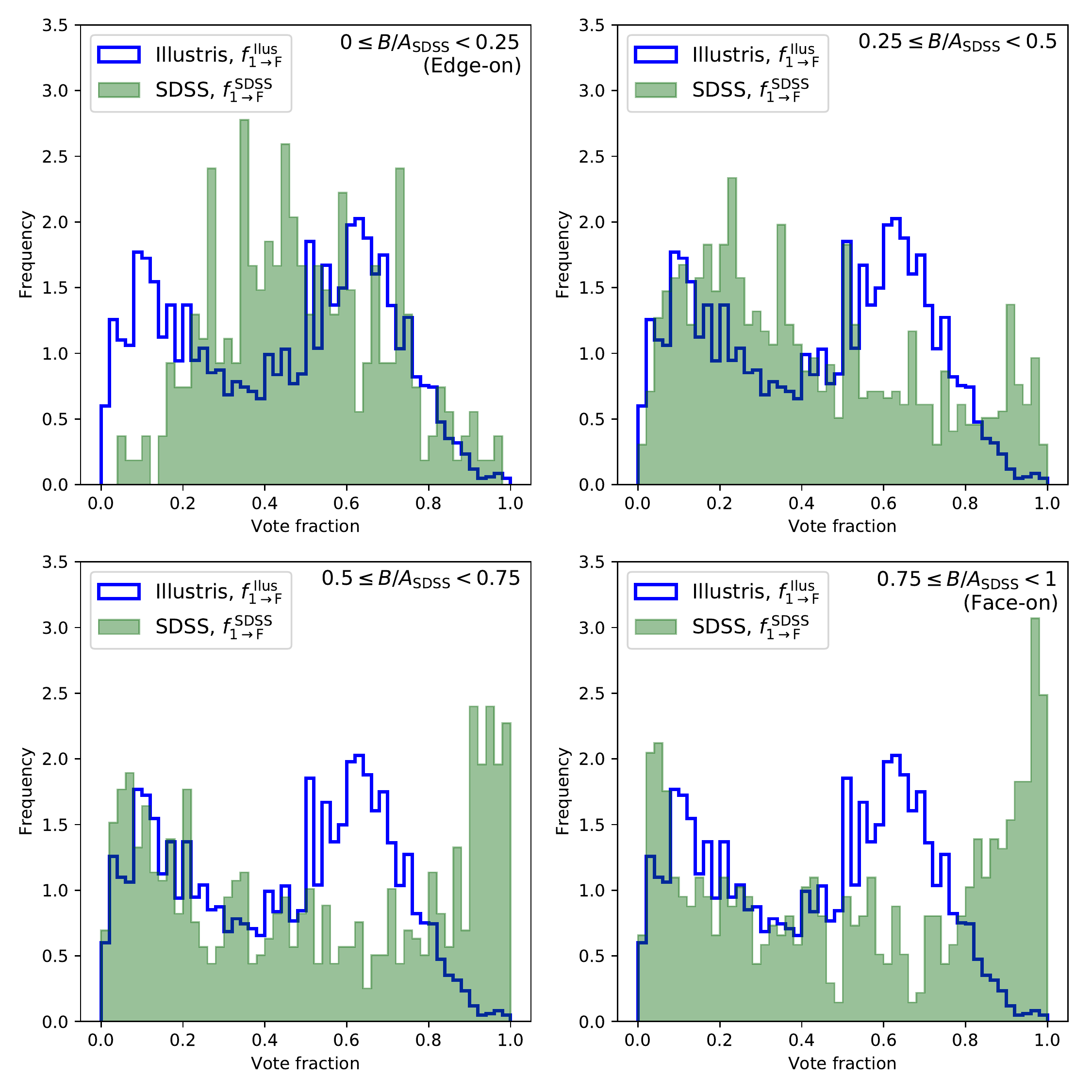}
  \caption{Distributions of $f_{\rm 1\rightarrow F}^{\mathrm{SDSS}}$ for disjoint subsets of the \sdss\ sample (\textit{green}) that were segregated based upon the observed axial ratio $(B/A)_{\mathrm{SDSS}}$ between the projected semiminor ($B$) and semimajor ($A$) axes of each galaxy. Proper interpretation of $f_{\rm 1\rightarrow F}$ is explained in the main text as well as in the caption of Figure \ref{fig:featuresVoteFracComparison}. Many more galaxies exhibit visible features (attain high $f_{\rm 1\rightarrow F}^{\mathrm{SDSS}}$) as $(B/A)_{\mathrm{SDSS}}$ increases from zero to unity. For comparison, the distribution of $f_{\rm 1\rightarrow F}^{\rm Illus}$ ({blue line}) for the entire \illus\ sample is shown in {all} panels.). For the \sdss\ distribution shown in the {upper-left} panel (smallest $(B/A)_{\mathrm{SDSS}}$, most edge-on), no galaxies were unambiguously classified as featured or smooth. This indicates that edge-on galaxies may be particularly difficult to separate base on visual inspection.}
  \label{fig:axialRatioSplitDists}
 \end{center}
\end{figure*}

The intentional omission of dust modeling when generating the synthetic \illus\ images (see $\S$\ref{subsec:remaining_inconsistencies}) is another factor that likely contributes to the mismatched visual classifications. To illustrate how intrinsic dust extinction affects the classifications that are gathered for \textit{real} galaxy images, Figure \ref{fig:axialRatioSplitDists} plots featured vote fraction distributions for disjoint subsets of the \sdss\ sample that were segregated based upon the observed axial ratio $(B/A)_{\mathrm{SDSS}}$ between the projected semi-minor ($B$) and semi-major ($A$) axes of each galaxy\footnote{The values for $A$ and $B$ correspond to those listed in the \sdss\ DR7 calatog for the \textit{exponential} or de Vaucouleurs profile model that provided the best fit to each galaxy's light distribution.}. Remarkable differences between the four distributions are evident with volunteers labeling many more featured galaxies as the typical axial ratio for each subset increases from zero (edge-on) to unity (face-on).

This phenomenon is likely dual in origin. Intrinsic dust extinction within the target galaxy may obscure discernible features while superimposed substructures along the line of sight may lead them to appear as a single luminous mass. Focusing on structurally disk-like galaxies, small values of $(B/A)_{\mathrm{SDSS}}$ suggest that the target was observed with an edge-on orientation. This configuration increases the probability of discrete substructures occupying nearby sightlines and becoming visually indistinguishable. Moreover, escaping starlight that would reveal such features must traverse a much larger column of dust on average without being absorbed in order to reach the observer. Conversely, as $(B/A)_{\mathrm{SDSS}}\rightarrow1$, galaxies with face-on orientations predominate and discrete substructures become more visible.

The procedure used to generate the \illus\ subject images did not model dust extinction, and we show the normalized featured vote fraction distribution for the full \illus\ sample in all four panels of Figure \ref{fig:axialRatioSplitDists}. The \illus\ and \sdss\ distributions do not coincide well for \textit{any} of the $(B/A)_{\mathrm{SDSS}}$ ranges considered. For $(B/A)_{\mathrm{SDSS}} \gtrsim 0.25$, the disparity is clearly manifested as an excess of apparently featured galaxies among the \illus\ sample. It is plausible that the galaxies contributing to this excess would shift to lower $f_{\rm 1\rightarrow F}$ if dust attenuation were properly simulated when preparing the \illus\ subject images. Such migration might dilute or even eliminate the apparent morphological disparities between the two samples.

\section{Summary and Conclusions}
We have used visual classifications from Galaxy Zoo to compare the coarse morphological appearance of simulated galaxies from the \illus\ cosmological simulation with  those of a population drawn from the \sdssfull, matched in mass and redshift. This set of visual classifications  allows a direct comparison to be made with observations, with any differences indicating potentially missing physics in the simulation, the inevitably limited resolution of such simulations, or the choices made in producing `observationally realistic' images. In any case, understanding how selection by morphology might influence comparisons between simulation and observation is essential.

Figure \ref{fig:featuresVoteFracComparison} reveals two marked disparities between the two samples. The fraction ($f_{\rm 1\rightarrow F}^{\rm Illus}$) of classifiers who report noticeable features in \illus\ galaxy images exceeds that for the equivalent quantity ($f_{\rm 1\rightarrow F}^{\rm SDSS}$) for classifications of \sdss\ subjects. Indeed, Figure \ref{fig:featuresVoteFracMassIntervalComparison} illustrates that for $\log(M_{\star}/M_{\odot}) < 10.5$ the distributions of $f_{\rm 1\rightarrow F}^{\rm Illus}$
and
$f_{\rm 1\rightarrow F}^{\rm SDSS}$ are almost mirror images of each other. While volunteer classifiers clearly discern features in a large majority of the \illus\ sample, a far smaller proportion report them for the \sdss\ galaxy images.  There is also a small set of galaxies with high featured vote fractions in \sdss\ but this population is absent in \illus. While the \illus\ images are simulated to an observational resolution of $1\arcsec$ compared to an achieved average seeing of $1.4\arcsec$ for the \sdss, this small difference is unlikely to be responsible for such a large observed difference.

The absence of moderate and high-mass, unambiguously featured galaxies in the \illus\ sample that was noted in $\S$\ref{sec:results} is perhaps the most surprising result. It may represent the response of volunteer classifiers to simulated objects, which, despite the care taken in preparing the images, are often easily distinguished from their \sdss\ counterparts. Features such as bright knots, over-prominent arms, and so on are seen in many \illus\ images. These artifacts are the result of insufficient particle resolution and  may confuse classifiers, reducing the consensus on features. Alternatively, it may be that the simulation is failing to producing realistic grand design spirals.

We also see a failure to produce the correct fraction of smooth galaxies. The importance of this mismatch between the \illus\ and \sdss\ samples appears to depend strongly on the stellar mass range of the galaxies under consideration. Figure \ref{fig:featuresVoteFracMassIntervalComparison} plots analogues of Figure \ref{fig:featuresVoteFracComparison} for {mass-selected} subsets of the \illus\ and \sdss. It is apparent that the distributions of $f_{\rm 1\rightarrow F}^{\rm Illus}$ and $f_{\rm 1\rightarrow F}^{\rm SDSS}$ become markedly less disparate for \textbf{stellar} masses $M_{\star} > 10^{11}M_{\odot}$. However, correspondence between the two datasets remains imperfect, and a population of highly featured galaxies that are present in the real universe, but absent in \illus\ becomes apparent above $M_{\star} > 10^{10.5} M_{\odot}$.

The underproduction of unambiguously featured galaxies with large $M_{\star}$ that we identify in \illus\ may indicate that accumulation of stellar mass involves simulated processes that also disrupt or destroy spatially discrete substructures. The most massive galaxies in \illus\ are predominantly formed by the hierarchical assembly of smaller systems \citep{2016MNRAS.458.2371R}. Repeated interactions between simulated galaxies provide a plausible mechanism for suppression of visible features. To investigate this possibility, we searched for indications that the time since the most recent major merging event in a simulated galaxy's history predicts its morphological classification for galaxies with $M_{\star} > 10^{11} M_{\odot}$. No compelling correlations were observed. The two-sample Kolmogorov--Smirnoff test yields a $p$-value of 0.104 when comparing the distributions of the time since the most recent major merging event for subsamples of visually smooth ($f_{\rm 1\rightarrow F}^{\rm Illus} < 0.3$) and featured ($f_{\rm 1\rightarrow F}^{\rm Illus} > 0.85$) galaxies. This is consistent with both subsamples being drawn from the same parent distribution. We also checked for a significant correlation between the fraction of galactic stellar mass that was formed in-situ and the visibility of features in the \illus\ galaxy images. In this case, the two-sample Kolmogorov--Smirnoff test yields a $p$-value of $4.1\times10^{-7}$ when comparing the samples of smooth and featured galaxies. This result indicates that $f_{\rm 1\rightarrow F}^{\rm Illus} \geq 0.85$ comprise a larger proportion of stars that were formed in-situ, which is broadly supportive of the hypothesis that visually featured galaxies experienced comparatively fewer interactions during their formation. A more rigorous verification that accumulation of ex-situ stellar mass is indeed responsible for the disruption of visually apparent substructures would require detailed examination of each galaxy's assembly history, which is beyond the scope of this paper.

Given that the ability of a simulation to represent a galaxy depends coarsely on the number of particles used to model it, some mass dependence should be expected; indeed, this is why galaxies with stellar masses less than $10^{10} M_{\odot}$ were excluded from the study. Such differences have been seen before, in particular by \citet{2017MNRAS.tmp...30B} who showed that a threshold at $M_{\star} > 10^{11} M_{\odot}$ also emerges when attempting morphological classification using parametric fits to the galaxy's light profile. Below this critical mass, the simulation produces a large proportion of disk-dominated galaxies; we confirm this result and show that it has a significant effect not only on the parametric measurements but on the overall visual morphology of the system being studied. \revision{}{In some cases, non-parametric morphological metrics for \illus\ galaxies also appear to differ from those of their physical counterparts when $M_{\star} \lesssim 10^{11} M_{\odot}$. For example, \citet{2017MNRAS.465.1106B} show that the measured asymmetry of merging \illus\ galaxies appears artificially large in comparison with mass-matched observational samples.}{dickinson} \revision{}{In the same mass range, \citet{2015MNRAS.454.1886S} identify a peculiar population of galaxies that exhibit distinctive ring-like structures of enhanced star formation, resulting in  unexpectedly extended morphologies (examples of several such systems are included in Figure \ref{fig:sdssIllustrisSubjectComparison}). \citet{2015MNRAS.454.1886S} suggest that these ring-like structures may reflect an imperfect model for coupling between feedback mechanisms and the interstellar medium (ISM) in \illus\ galaxies. Alternatively, the rings of star formation may be an inherent manifestation of the ISM equation of state that is assumed for the \illus\ simulation.}{dickinson} \revision{}{Earlier studies \citep[e.g.][]{2011MNRAS.418..801H} compared the properties of simulated galaxy samples with those of locally observed systems using non-parametric morphological estimators. Similar discrepancies pertaining to excessive asymmetry and clumpy substructure were identified.}{dickinson}

As in \illus\ a galaxy's stellar mass broadly maps to the number of stellar particles comprising the simulated galaxy, we conclude that below $10^{11} M_{\odot}$, the number of stellar particles comprising a galaxy is apparently insufficient to represent the simulated physics reliably, and observed structures are often likely to result from resolution-induced artifacts. The effects are subtle, and the images produced by the simulation are clearly perceived as realistic, but as a population there remain differences between simulated and observed galaxies. \revision{}{These differences complicate more detailed comparisons between the \illus\ and \sdss\ galaxy morphologies. Below $M_{\star}\sim10^{11}M_{\odot}$, the coarse morphological differences between observed and simulated galaxies could artificially distort the later stages of classification, because early volunteer responses restrict the set of questions that are subsequently posed. For the most massive galaxies, a limited number of subject images results in excessively sparse sampling of the Galaxy Zoo classification hierarchy that prevents reliable inference of morphological characteristics.}{dickinson} Future studies that match \sdss\ and \illus\ samples should be aware of \revision{this threshold}{the $10^{11} M_{\odot}$ threshold we have identified}{} and its effects on the comparison being made. We have also shown that insight can be derived from visual analysis of large samples of images derived from simulations and recommend this procedure for future data products.

\acknowledgements
\emph{Acknowledgements:} The data in this paper are the result of the efforts of the Galaxy Zoo volunteers, without whom none of this work would be possible. Their efforts are individually acknowledged at \url{authors.galaxyzoo.org}. Please contact the author(s) to request access to research materials discussed in this paper.

H.D., L.F., C.S., M.B., and M.G. gratefully acknowledge support from the US National Science Foundation grant AST1716602 (H.D., L.F., C.S. also supported by NSF grant AST1716602).

C.J.L. was supported by STFC under grant ST/N003179/1.

B.D.S. acknowledges support from the National Aeronautics and Space Administration (NASA) through Einstein Postdoctoral Fellowship Award Number PF5-160143 issued by the Chandra X-ray Observatory Center, which is operated by the Smithsonian Astrophysical Observatory for and on behalf of NASA under contract NAS8-03060.

\revision{}{P.T. acknowledges support from NASA through Hubble Fellowship grants HST-HF2-51384.001-A awarded by the STScI, which is operated by the Association of Universities for Research in Astronomy, Inc., for NASA, under contract NAS5-26555.}{dickinson}

Funding for the SDSS and SDSS-II has been provided by the Alfred P. Sloan Foundation, the Participating Institutions, the National Science Foundation, the U.S. Department of Energy, the National Aeronautics and Space Administration, the Japanese Monbukagakusho, the Max Planck Society, and the Higher Education Funding Council for England. The SDSS website is \url{http://www.sdss.org/}.
The SDSS is managed by the Astrophysical Research Consortium for the Participating Institutions. The Participating Institutions are the American Museum of Natural History, Astrophysical Institute Potsdam, University of Basel, University of Cambridge, Case Western Reserve University, University of Chicago, Drexel University, Fermilab, the Institute for Advanced Study, the Japan Participation Group, Johns Hopkins University, the Joint Institute for Nuclear Astrophysics, the Kavli Institute for Particle Astrophysics and Cosmology, the Korean Scientist Group, the Chinese Academy of Sciences (LAMOST), Los Alamos National Laboratory, the Max-Planck-Institute for Astronomy (MPIA), the Max-Planck-Institute for Astrophysics (MPA), New Mexico State University, Ohio State University, University of Pittsburgh, University of Portsmouth, Princeton University, the United States Naval Observatory, and the University of Washington.

\bibliographystyle{aasjournal}
\bibliography{IllustrisGZ}

\end{document}